\documentclass[manuscript,screen,nonacm,format=sigconf]{acmart}

\usepackage{booktabs}
\usepackage{amsmath}
\usepackage{tikz}
\usepackage{algorithm}
\usepackage{algpseudocode}
\usetikzlibrary{positioning,fit,backgrounds,arrows.meta}

\usepackage{booktabs}
\usepackage{graphicx}
\usepackage[table]{xcolor}
\usepackage{array}

\definecolor{tsblue}{HTML}{EAF2FF}
\definecolor{lobteal}{HTML}{E8F7F6}
\definecolor{tabgold}{HTML}{FFF6DF}
\definecolor{structviolet}{HTML}{F2ECFF}

\newcommand{\TS}{\cellcolor{tsblue}\textbf{Time series}}
\newcommand{\LOB}{\cellcolor{lobteal}\textbf{LOB}}
\newcommand{\TAB}{\cellcolor{tabgold}\textbf{Tabular}}
\newcommand{\STR}{\cellcolor{structviolet}\textbf{Structured}}

\definecolor{finBlue}{HTML}{2F6F9F}
\definecolor{finTeal}{HTML}{2A9D8F}
\definecolor{finGold}{HTML}{C49A27}
\definecolor{finRed}{HTML}{B55A4A}
\definecolor{finViolet}{HTML}{6D5BA6}
\definecolor{finInk}{HTML}{263238}
\definecolor{finMist}{HTML}{EEF3F6}

\begin{document}

\title{Diffusion Models in Finance: A Survey}

\author{Zhuohan Wang}
\affiliation{%
  \institution{King's College London}
  \city{London}
  \country{United Kingdom}}
\email{zhuohan.wang@kcl.ac.uk}

\author{Carmine Ventre}
\affiliation{%
  \institution{King's College London}
  \city{London}
  \country{United Kingdom}}
\email{carmine.ventre@kcl.ac.uk}

\renewcommand{\shortauthors}{Anonymous Author(s)}

\begin{abstract}

Diffusion generative models have rapidly emerged as powerful tools for modeling complex financial data. Their appeal is both structural and practical: they offer stable likelihood-based training, strong mode coverage, flexible conditioning, and a stochastic-differential-equation formulation that aligns naturally with the It\^o calculus and stochastic control frameworks widely used in finance. This survey reviews the growing literature on diffusion-family generative models for financial applications. We organize prior work primarily by financial data type, covering time series, limit order books, tabular data, and other structured financial objects, while discussing the modeling goals and application contexts that arise within each category. To the best of our knowledge, this is the first survey dedicated specifically to diffusion-family models for financial data. For more detailed information, we have open-sourced a repository\footnote{\url{https://github.com/ZhuoHan1998/Diffusion-Models-In-Finance}}. 
\end{abstract}


\maketitle

\section{Introduction}\label{sec:intro}

Financial markets play a central role in modern economies by allocating capital, transferring risk, and supporting investment and consumption decisions. However, financial data are notoriously difficult to model, which often exhibit heavy tails, volatility clustering, nonlinear dependence, regime changes, and abrupt responses to macroeconomic or geopolitical events~\cite{bollerslev1986garch,heston1993sv,gatheral2018rough}. These properties make it challenging for traditional parametric models to fully capture the distributional complexity of financial markets, especially when the goal is to support decision under uncertainty. 
Generative modeling has therefore become increasingly relevant in finance. Rather than estimating a single expected outcome, generative models aim to learn the underlying data distribution and produce plausible samples from it. This is particularly useful for applications such as financial data generation, stress testing, portfolio risk analysis, market simulation, derivative pricing, etc.~\cite{potluru2023synthetic, wiese2020quantgan,ni2021sigwgan,sattarov2023findiff}.

Diffusion models have recently emerged as one of the most powerful classes of generative models. Originally popularized in computer vision and later extended to domains such as language, time series and tabular data~\cite{yang2023diffusion}, diffusion models generate samples through a gradual denoising process~\cite{sohldickstein2015,ho2020ddpm,song2021sde}. Compared with earlier generative approaches such as generative adversarial networks~\cite{goodfellow2020generative}, variational autoencoders~\cite{kingma2013auto}, and normalizing flows~\cite{rezende2015variational}, diffusion models offer attractive properties including stable training, high sample quality, flexible conditioning mechanisms, and strong empirical performance in modeling complex distributions~\cite{dhariwal2021beatgan,nichol2021improved,ho2022cfg,karras2022edm}.
These characteristics make diffusion models a promising candidate for financial applications. Financial data are noisy, high-dimensional, non-stationary, and often governed by latent market regimes. A generative framework that can learn rich distributions and condition on market states, historical information, or external variables may therefore be useful for generating synthetic price paths, estimating risk measures, improving forecasting systems, and supporting downstream decision-making~\cite{tanaka2025cofindiff,takahashi2024synthfts,chen2025difffactor,cho2025diffolio}. 


However, the use of diffusion models in finance remains relatively new, and the literature is dispersed across applications including time-series generation, limit order books, tabular data, and structured financial objects~\cite{sattarov2023findiff,berti2025trades,jin2025ivsurface,zhao2025exotic}. 
Existing surveys cover synthetic data in finance~\cite{potluru2023synthetic}, diffusion models for generic time series~\cite{yang2024tssurvey}, diffusion models for tabular data~\cite{li2025diffusion}, and large language models or agents in finance~\cite{li2023large, saha2025large}, but none provides a dedicated overview of diffusion models across financial applications.

\begin{figure*}[t]
  \centering
  \includegraphics[width=0.93\textwidth,height=0.37\textheight]{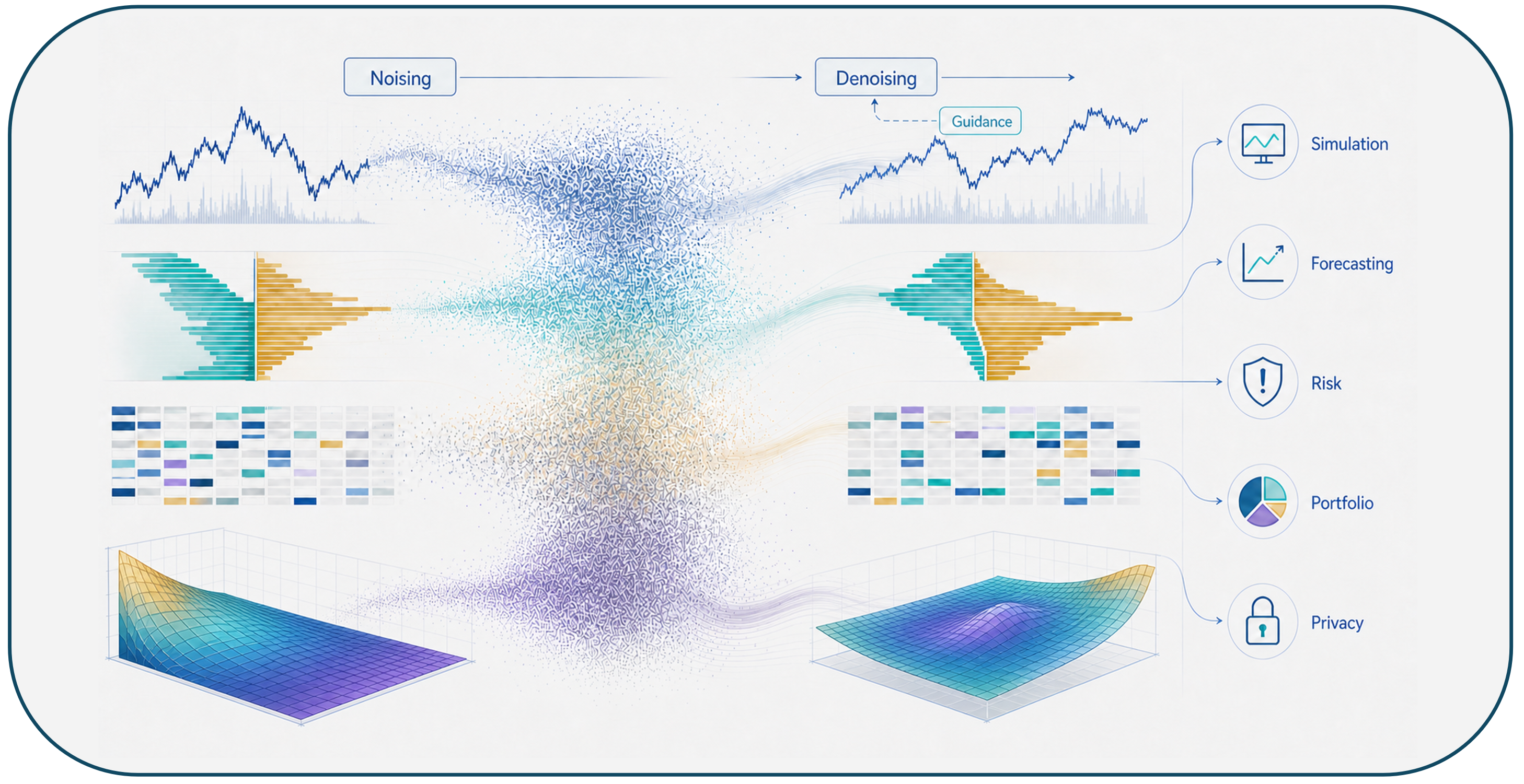}
  \caption{Taxonomy of diffusion-family models in finance organized by financial data type. Data are generated through noising and guided denoising, then used for downstream tasks such as simulation, forecasting, risk analysis, portfolio construction, and privacy preservation.}
  \label{fig:taxonomy}
\end{figure*}

This survey provides a structured overview of diffusion models in finance by reviewing methodological foundations, organizing applications across financial time series, limit order books, tabular data, and other structured financial objects, and identifying future research directions. The remainder of the paper is organized as follows: Section~\ref{sec:bg} introduces diffusion models, flow models, and related generative modeling foundations; Section~\ref{sec:fts} to Section~\ref{sec:structured} review applications across the main financial data modalities; and Section~\ref{sec:conclusion} discusses open problems and concluding remarks. Figure~\ref{fig:taxonomy} illustrates the conceptual pipeline behind the survey. 

\section{Background}\label{sec:bg}
This section fixes the common vocabulary used throughout the survey. We keep the presentation deliberately compact: the goal is not to re-derive diffusion models, but to identify the design choices that become consequential once the data are time series, limit order books, tabular data, or other financial objects.

\textbf{Diffusion-based models.}
DDPMs define a forward noising process that gradually transforms data $x_0 \sim p_{\rm data}$ into an approximately Gaussian variable $x_T$, typically through transitions $q(x_t \mid x_{t-1})$ with a prescribed noise schedule~\cite{ho2020ddpm}. A neural network is then trained to reverse this process, often by predicting the injected noise. The practical appeal is that training is stable and uses a simple denoising objective, while sampling produces diverse draws by running the learned reverse chain from noise to data. Score-based generative models instead learn $s_\theta(x,t,c)\approx \nabla_x \log p_t(x\mid c)$ at multiple noise levels~\cite{song2019ncsn}. The two views meet in the score-SDE formulation~\cite{song2021sde}, where a forward process
\begin{equation}
  d x_t = f_t(x_t)\,dt + g_t\,d w_t
\end{equation}
is sampled in reverse by replacing the unknown score with a neural estimate:
\begin{equation}
  d x_t = \{f_t(x_t)-g_t^2 s_\theta(x_t,t,c)\}\,dt + g_t\,d\bar{w}_t.
\end{equation}
Here $c$ denotes optional conditioning information and $\bar{w}_t$ is the reverse-time Brownian motion. Training usually samples $x_0$, a noise level $t$, and a perturbed observation $x_t=\alpha_t x_0+\sigma_t\epsilon$, then fits either the score or the injected noise. Algorithm~\ref{alg:sde} summarizes this diffusion-based training and Euler--Maruyama sampling. This formulation is especially natural in finance because stochastic differential equations are already widely used to describe price dynamics. However, instead of calibrating a fixed drift or volatility coefficient for an assumed market process like a classical financial diffusion model, the neural network learns the reverse-time score, or equivalently a denoising drift, that transforms simple noise into samples from the empirical financial data distribution.

\begin{algorithm}[t]
\caption{Diffusion-based models: training and sampling}
\label{alg:sde}
\small
\begin{algorithmic}[1]
\State \textbf{Input:} A dataset of samples $x_0 \sim p_{data}$, optional condition $c$, schedule $(\alpha_t,\beta_t)$, score network $s_{t}^{\theta}$ or noise predictor $\epsilon_t^{\theta}$.
\Statex \textit{Training}
\For{each minibatch}
  \State Sample $t\sim U(0,T)$ and $\epsilon\sim\mathcal{N}(0,I)$.
  \State Set $x_t=\alpha_t x_0+\beta_t\epsilon$.
  \State Minimize $\|s_t^{\theta}(x_t,c)-\nabla\log p_t(x_t\mid x_0)\|^2$.
  \State Equivalently, under Gaussian paths, minimize $\|\epsilon_t^{\theta}(x_t,c)-\epsilon\|^2$.
\EndFor
\Statex \textit{Sampling}
\State Draw $x_T\sim\mathcal{N}(0,I)$.
\For{$t=T,T-\Delta t,\ldots,\Delta t$}
  \State Draw $z\sim\mathcal{N}(0,I)$.
  \State $x_{t-\Delta t}=x_t-\{f_t(x_t)-g_t^2 s_\theta(x_t,c)\}\Delta t+g_t\sqrt{\Delta t}\,z$.
\EndFor
\State \Return synthetic sample $\hat{x}_0$.
\end{algorithmic}
\end{algorithm}

\textbf{Flow-based models.}
Recent work also uses continuous-time transport formulations that learn a vector field from a base distribution to the data distribution. In its simplest form, flow matching
learns an ODE
\begin{equation}
  \frac{d x_t}{dt}=v_\theta(x_t,t,c), \qquad x_0\sim p_0,
\end{equation}
whose terminal distribution approximates $p_{\rm data}$. With a linear conditional path $x_t=(1-t)z+t x$ between noise $z\sim p_0$ and data $x\sim p_{\rm data}$, the target velocity is $u_t=x-z$, giving the objective
\begin{equation}
  \mathcal{L}_{\rm FM}(\theta)=
  \mathbb{E}_{x,z,t}\left[\|v_\theta(x_t,t,c)-(x-z)\|_2^2\right].
\end{equation}
Algorithm~\ref{alg:flow} summarizes the corresponding training and Euler sampling~\cite{lipman2023flow,holderrieth2026introflowdiffusion}.
Diffusion model families are increasingly understood as instances of a unified generative-transport view. Stochastic interpolants unify flows and diffusions by learning dynamics along paths between a simple base distribution and the data distribution~\cite{albergo2023interpolants}, while recent tutorials on flow matching and diffusion models emphasize that both learn fields that transport noise into data under different sampling dynamics and training targets
~\cite{holderrieth2026introflowdiffusion}.

\begin{algorithm}[t]
\caption{Flow-based models: training and sampling}
\label{alg:flow}
\small
\begin{algorithmic}[1]
\State \textbf{Input:} data $x$, base noise $z\sim p_0$, optional condition $c$, vector field $v_\theta$.
\Statex \textit{Training}
\For{each minibatch}
  \State Sample $t\sim U(0,1)$, $z\sim p_0$, and $x\sim p_{\rm data}$.
  \State Set $x_t=(1-t)z+t x$ and target velocity $u_t=x-z$.
  \State Minimize $\|v_\theta(x_t,t,c)-u_t\|_2^2$.
\EndFor
\Statex \textit{Sampling}
\State Draw $x_0\sim p_0$.
\For{$t=0,\Delta t,\ldots,1-\Delta t$}
  \State $x_{t+\Delta t}=x_t+\Delta t\,v_\theta(x_t,t,c)$.
\EndFor
\State \Return synthetic sample $\hat{x}_1$.
\end{algorithmic}
\end{algorithm}

\textbf{Guidance.}
Classifier guidance modifies the reverse process using gradients from an auxiliary model~\cite{dhariwal2021beatgan}, while classifier-free guidance learns conditional and unconditional scores in a single model and interpolates them at sampling time~\cite{ho2022cfg}. Conditioning layers such as FiLM provide a general mechanism for injecting external variables by feature-wise modulation~\cite{perez2018film}. Finance papers adapt this idea to regime labels~\cite{tanaka2025cofindiff,wang2026difflob}, factor states~\cite{gao2024diffsformer,gao2025factorcond}, asset identities~\cite{daiya2024diffstock}, stress scenarios~\cite{choudhary2025diffaugrl}, portfolio objectives~\cite{gao2025factorcond,alzahrani2025marcd}, and structural constraints~\cite{kubiak2024corrddpm,sattarov2024imbfindiff}.

\textbf{Sampling accelerators.}
Sampling speed is a recurring bottleneck, especially in finance where simulation, stress testing, and portfolio evaluation may require many generated paths. DDIM~\cite{songddim2021}, improved diffusion objectives~\cite{nichol2021improved}, EDM design choices~\cite{karras2022edm}, and consistency models~\cite{song2023consistency} improve sample quality per function evaluation. Training-free numerical solvers further accelerate sampling by treating generation as solving a diffusion ODE, including DPM-Solver~\cite{lu2022dpm}, DPM-Solver++~\cite{lu2025dpm}, and unified predictor-corrector solvers such as UniPC~\cite{zhao2023unipc}.

\textbf{Architectural backbones.}
Backbone choice determines what structure a diffusion-family model can represent. TimeGrad supports probabilistic time-series forecasting~\cite{rasul2021timegrad}, CSDI targets conditional imputation~\cite{tashiro2021csdi}, SSSD combines diffusion with structured state-space sequence modeling~\cite{alcaraz2022sssd}, and TSDiff/Diffusion-TS study forecasting-oriented and interpretable time-series generation~\cite{kollovieh2023tsdiff,yuan2024diffusionts}. For tabular settings, TabDDPM and TabSyn adapt diffusion to mixed-type data~\cite{kotelnikov2023tabddpm,zhang2024tabsyn}, while transformer-based denoisers and rectified-flow or stochastic-interpolant models broaden the design space beyond standard U-Net-style diffusion~\cite{peebles2022dit,liu2023rectifiedflow,albergo2023interpolants,hu2024flowts}. Finance papers typically do not adopt these backbones unchanged; rather, they adapt the same design principles to specific financial structure. 

\begin{table*}[t]
\centering
\caption{Representative diffusion-family finance papers organized by data domain. 
The table is illustrative rather than exhaustive; entries are selected to span the
main data objects, applications, and model families used in this survey.}
\label{tab:representative}
\scriptsize
\setlength{\tabcolsep}{2pt}
\renewcommand{\arraystretch}{1.08}
\resizebox{\textwidth}{!}{
\begin{tabular}{@{}
p{0.095\textwidth}
p{0.235\textwidth}
p{0.045\textwidth}
p{0.16\textwidth}
p{0.08\textwidth}
p{0.055\textwidth}
p{0.075\textwidth}
p{0.125\textwidth}
p{0.16\textwidth}
@{}}
\toprule
Data domain & Paper & Year & Venue & Model & Region & Asset & Generated object & Application \\
\midrule

\TS & DiffSTOCK~\cite{daiya2024diffstock} & 2024 & ICASSP & Diffusion & US & Stock & Price series & Prediction / portfolio \\
\TS & DiffsFormer~\cite{gao2024diffsformer} & 2024 & arXiv & Diffusion & CN & Stock & Factor series & Factor augmentation \\
\TS & Beyond Monte Carlo~\cite{lesniewski2024beyondmc} & 2024 & arXiv & Diffusion & US & Stock & Return vectors & Market simulation \\
\TS & Financial Time Series Denoiser~\cite{wang2024ftsdenoiser} & 2024 & ICAIF & Diffusion & US & Stock & Price series & Denoising / trading signal \\
\TS & Generation of Synthetic Financial Time Series~\cite{takahashi2024synthfts} & 2025 & Quantitative Finance & Diffusion & US & Stock & Return and volume series & Synthetic generation \\
\TS & Diffusion Factor Models~\cite{chen2025difffactor} & 2025 & arXiv & Diffusion & US & Stock & Return series & Factor-structured generation \\
\TS & CoFinDiff~\cite{tanaka2025cofindiff} & 2025 & IJCAI & Diffusion & JP & Stock & Price series & Controllable generation \\
\TS & Diffusion-Augmented RL~\cite{choudhary2025diffaugrl} & 2025 & arXiv & Diffusion & US & Stock & Return series & Stress scenarios / portfolio \\
\TS & Factor-Based Conditional Diffusion~\cite{gao2025factorcond} & 2025 & arXiv & Diffusion & CN & Stock & Return series & Portfolio optimization \\
\TS & FlowHFT~\cite{li2025flowhft} & 2025 & arXiv & Flow & Synthetic & Stock & Trading actions & HFT imitation policy \\
\TS & Diffolio~\cite{cho2025diffolio} & 2026 & Information Fusion & Diffusion & US & Stock & Return series & Forecasting / portfolio \\

\midrule

\LOB & TRADES~\cite{berti2025trades} & 2025 & ECAI & Diffusion & US & Stock & LOB orderflow & Market simulation \\
\LOB & DiffVolume~\cite{wang2025diffvolume} & 2025 & ICAIF & Diffusion & US & Stock & LOB volume & Volume generation \\
\LOB & Painting the Market~\cite{backhouse2025painting} & 2025 & arXiv & Diffusion & US & Stock & LOB snapshots & Simulation / forecasting \\
\LOB & Controllable Financial Market Generation~\cite{huang2024digma} & 2026 & AAAI & Diffusion & CN & Stock & LOB orderflow & Controllable market generation \\
\LOB & Limit Order Book Event Stream Prediction~\cite{zheng2024lobdif} & 2026 & Data Science and Engineering & Diffusion & US, CN & Stock & LOB orderflow & Event-stream prediction \\
\LOB & DiffLOB~\cite{wang2026difflob} & 2026 & IJCAI-ECAI & Diffusion & US & Stock & LOB snapshots & Counterfactual generation \\

\midrule

\TAB & FinDiff~\cite{sattarov2023findiff} & 2023 & ICAIF & Diffusion & Multi & Multi & Tabular records & Tabular synthesis \\
\TAB & Entity-based Financial Tabular Synthesis~\cite{liu2024enttabdiff} & 2024 & ICAIF & Diffusion & Multi & Multi & Tabular records & Entity-based synthesis \\
\TAB & Imb-FinDiff~\cite{sattarov2024imbfindiff} & 2024 & ICAIF & Diffusion & Multi & Multi & Tabular records & Imbalance synthesis \\
\TAB & Latent-Space Flow-Based Diffusion~\cite{ihsan2025latentflow} & 2025 & arXiv & Flow & Multi & Multi & Tabular records & Data augmentation \\
\TAB & DP-FinDiff~\cite{sattarov2025dpfindiff} & 2025 & arXiv & Diffusion & Multi & Multi & Mixed-type tabular & Private synthesis \\
\TAB & EmDT~\cite{kuo2026emdt} & 2026 & arXiv & Diffusion & Multi & Fraud & Tabular records & Fraud augmentation \\
\TAB & Privacy Risks and Tradeoffs~\cite{zuo2026privacyrisks} & 2026 & arXiv & Diffusion & Multi & Multi & Tabular records & Privacy evaluation \\

\midrule

\STR & Correlation Matrix Diffusion~\cite{kubiak2024corrddpm} & 2024 & ICAIF & Diffusion & US & Stock & Correlation matrix & Correlation synthesis \\
\STR & IV-Surface Diffusion~\cite{jin2025ivsurface} & 2025 & arXiv & Diffusion & US & Option & IV surface & Surface forecasting \\
\STR & Exotic Options and Counterparty Games~\cite{zhao2025exotic} & 2025 & arXiv & Diffusion & Global & Option & Option price paths & Exotic option valuation \\
\STR & Term-Structure Diffusion~\cite{fukunishi2025termstructure} & 2026 & Expert Systems with Applications & Diffusion & JP & Interest rate & Yield curve & Term-structure generation \\
\STR & Risk-Neutral Derivative Pricing~\cite{tiwari2026riskneutral} & 2026 & arXiv & Diffusion & Global & Stock & Risk-neutral asset paths & Derivative pricing \\
\STR & TF-CoDiT~\cite{zhang2026tfcodit} & 2026 & arXiv & Diffusion & CN & Treasury future & Price series & Fixed-income synthesis \\

\bottomrule
\end{tabular}}
\end{table*}


\section{Time Series Data}\label{sec:fts}

Financial time series form the largest cluster in the current literature. In this section, we use the term narrowly: the modeled object is an ordered sequence such as prices, returns, macroeconomic variables, etc. We therefore separate this section from LOBs in section~\ref{sec:lob}, where the sequence is constrained by order-book mechanics, and from structured objects such as implied-volatility surfaces or yield curves. The dominant assets are equities, with smaller coverage of ETFs, futures, energy prices, and macro-conditioned paths. Across these settings, the generator must preserve stylized facts that ordinary time-series metrics often miss, such as heavy tails, volatility clustering, leverage effects, serial dependence and cross-sectional covariance.

\subsection{Unconditional Generation}\label{sec:fts-uncond}

Unconditional generation asks whether a model can synthesize financial time series without specifying a future condition or control signal. We read these papers along two axes: the financial variable being generated and the dimensionality of the generated object. The variable axis separates price paths from returns, spreads, volumes, and transformed market features. The dimensionality axis separates single-instrument generation from multivariate panels that must preserve cross-asset dependence. 

Most unconditional finance papers in the current corpus are stock-market papers. \citet{takahashi2024synthfts} use US intraday stock data and generate a single-stock, multivariate-feature time series, where log returns, spreads, and trading volumes are converted to wavelet images and reconstructed as synthetic market series. \citet{lesniewski2024beyondmc} use US equity data to generate a multivariate stock-return scenario used for covariance regularization and portfolio-level simulation. Diffusion Factor Models~\cite{chen2025difffactor} move to high-dimensional stock-return panels, generating multivariate asset returns under a latent factor structure rather than independent univariate series. \citet{kim2025gbmdiff} use historical stock data and generate price-path-oriented series with a GBM-informed noising process, making the model closer to price dynamics than return-only scenario generation. \citet{liu2024synthetic} directly target synthetic asset-price-path generation with a DDPM, placing the work on the price-series side of unconditional financial time-series generation rather than the return-scenario side. \citet{masi2026high} propose a GAN--diffusion framework for high-quality synthetic financial time series, jointly generating stock mid-price and volume series while using the GAN critic to improve the realism of cross-asset correlation structures.  Fiaingen~\cite{rozanec2025fiaingen} is useful as a synthetic-data quality benchmark rather than merely a core diffusion architecture: it proposes graph-based generative methods for financial time series and evaluates whether synthetic stock-price windows resemble real data, generate quickly, and support downstream machine-learning tasks.

\subsection{Conditional Generation}\label{sec:fts-cond}
Conditional generation studies whether a model can synthesize financial time series under a user-specified condition. Compared with unconditional generation, the key question is not only what variable is generated, but also what information controls the generation. We read this literature along two axes: the generated data object, such as prices, returns or charts, and the conditioning signal, such as trend, volatility, partial observations, stress regimes, or optimization states. This distinction is important because conditional generators are usually evaluated by whether they respond correctly to the condition, not only by whether their samples look realistic on average.

For stock data, conditional generation mainly appears as controllable stock time-series synthesis and stress-scenario construction. CoFinDiff~\cite{tanaka2025cofindiff} uses Japanese stock price series and generates multivariate stock time series conditioned on user-specified trend and volatility controls, so the generated object is price-like stock series rather than return-only scenarios. 
InterDiff~\cite{long2025interdiff} conditions stock time-series synthesis on learned intra- and inter-stock correlation representations, using hierarchical transformers and classifier-free guidance to generate Chinese equity series that preserve cross-stock dependence.
\citet{choudhary2025diffaugrl} use a conditional DDPM to generate synthetic Dow-30 return sequences parameterized by crash intensity, augmenting PPO training with crisis-like market episodes so that the resulting portfolio policy is more robust to tail-risk and out-of-sample stress events. ~\citet{guo2025compressedsensing} uses financial time-series observations under partial or compressed measurements and reconstructs plausible stock and macro time-series signals, making the generated object a conditionally completed financial series rather than an unconditional market sample.

Outside stocks, \citet{alzahrani2025marcd} generate regime-conditioned ETF return scenarios with a tail-weighted diffusion model and use them in a turnover-constrained CVaR portfolio allocator. ~\citet{zarifis2025scenariotree} use multivariate energy price time series and generates conditional scenario paths for multistage stochastic optimization, and evaluate it on energy arbitrage in New York State’s day-ahead electricity market.

\subsection{Prediction and Trading}\label{sec:fts-decision}

Many forecasting and trading papers also use conditional diffusion models, so the distinction from Section~\ref{sec:fts-cond} is based on the model's primary purpose rather than its architecture. We discuss papers here when diffusion mainly supports prediction, portfolio allocation, or trading decisions, and reserve Section~\ref{sec:fts-cond} for work whose main contribution is controllable financial time-series synthesis.
Prediction-oriented papers use diffusion models to forecast future financial states or to construct signals for trading and allocation. In this subsection, prediction covers point and distributional forecasts of prices, returns, chart movements, denoised signals, portfolio weights, and trading actions. These models are usually evaluated by downstream decision value, including risk-adjusted returns, drawdown control, transaction-cost robustness, and performance across market regimes.

Stock-level prediction papers mostly use diffusion as a way to denoise, augment, or represent financial time-series signals. \citet{koa2023diffusion} propose a diffusion variational autoencoder for multi-step stock-price prediction, using diffusion to address stochasticity in future price trajectories. \citet{wang2024ftsdenoiser} apply a diffusion denoiser to US stock price series before downstream prediction and trading evaluation. \citet{fang2024spatio} use a spatio-temporal diffusion model to infer missing and real-time financial data in firm-characteristic panels. \citet{daiya2024diffstock} generate probabilistic stock-market predictions from relational stock time-series data, making the modeled object a cross-sectional prediction distribution rather than a standalone synthetic series. \citet{gao2024diffsformer} use Chinese stock factor and price-related data, with diffusion-based factor augmentation serving the prediction model rather than general market simulation. \citet{chen2025dhmoe} use diffusion-generated hierarchical multi-granular experts for stock prediction across US, Chinese, and Hong Kong markets. \citet{lee2025charts} recast financial forecasting as conditional image generation: historical price charts are treated as visual inputs, and a text-to-image diffusion model generates the next chart image to predict future price trends. \citet{nguyen2026vardiff} further combine visual representations with retrieval-guided diffusion for stock forecasting.

A second group uses diffusion-generated distributions not merely as forecasts, but as inputs to portfolio optimization and trading policies. \citet{cho2025diffolio} use Diffolio to forecast the joint distribution of future asset returns, combining hierarchical asset/market attention with a correlation-guided regularizer so that the generated forecasts are directly useful for portfolio construction. \citet{gao2025factorcond} condition stock-return generation on asset-specific factors and use the resulting samples for constrained mean-variance and mean-CVaR portfolio optimization. \citet{aghapour2025dynport} use score-based diffusion within dynamic portfolio selection, connecting generated return scenarios to sequential allocation decisions. \citet{bagchi2026factordim} evaluate conditional diffusion models for portfolio construction on large-scale equity data, showing that factor dimensionality controls a bias--variance tradeoff: too few factors underfit return dynamics, while too many factors overfit and produce unstable, concentrated portfolios. \citet{li2025flowhft} use flow matching for high-frequency market making, generating market-conditioned bid--ask action sequences from expert demonstrations rather than forecasting prices directly. 

\section{Limit Order Book Data}\label{sec:lob}

Limit order book (LOB) data describe the visible supply and demand around the best bid and ask. Unlike financial time series, LOB data are event-driven, high-dimensional, and constrained by market mechanics: prices must respect tick sizes, bid prices must remain below ask prices, queue sizes are nonnegative, and order flow must evolve in a temporally coherent way. These constraints make LOB generation more difficult than generic multivariate time-series synthesis. We organize this section by the generated object: orderflow generation models event streams, whereas orderbook generation models book states.

\subsection{Orderflow Generation}\label{sec:lob-orderflow}

Order-flow generation papers synthesize sequences of market events, including order submissions, cancellations or amends. \citet{berti2025trades} propose TRADES, a transformer-based denoising diffusion engine that generates market-state-conditioned LOB orderflow for realistic and responsive market simulation, with evaluation focused on synthetic-data usefulness and responsiveness to trading-agent interventions. ByteGen~\cite{li2025bytegen} works at the orderbook-event level with a tokenizer-free byte-space generator, making it useful context for event-stream generation even though its architecture is not a standard diffusion model. \citet{zheng2024lobdif} propose LOBDIF, which treats LOB event-stream prediction as learning a joint time--event distribution with a diffusion model, using a denoising network and skip-step sampling to forecast both the type and timing of future order-book events. \citet{huang2024digma} propose DigMA, which uses conditional diffusion to generate time-varying market-state parameters, such as mid-price return and order-arrival intensity, and then guides an economics-informed meta-agent to sample realistic and controllable order flow.

\subsection{Orderbook Generation}\label{sec:lob-orderbook}

Order-book generation papers synthesize structured book states, snapshots, or counterfactual LOB trajectories. DiffVolume \cite{wang2025diffvolume} focus on conditional generation of future LOB volume snapshots from past volume history and time-of-day information, with optional liquidity-profile conditioning for counterfactual volume scenarios. \citet{backhouse2025painting} convert LOB data into structured images and use diffusion inpainting to generate future book states in parallel, reducing autoregressive error accumulation while achieving strong LOB-Bench performance. \citet{wang2026difflob} propose DiffLOB for controllable counterfactual LOB generation, conditioning future book trajectories on hypothetical regimes such as trend, volatility, liquidity, and order-flow imbalance to support stress testing and scenario analysis.

\section{Tabular Data}\label{sec:tabular}

Financial tabular data consist of row-wise records such as credit applications, payment transactions, customer attributes, fund characteristics, fraud labels, and account-level features. Unlike financial time series or LOBs, these data are usually mixed-type: continuous fields, categorical variables, binary indicators, timestamps, labels, and entity identifiers may appear in the same table. The main modeling challenge is therefore not temporal coherence, but preserving feature-label relationships, categorical distributions, business constraints, rare classes, and privacy. We organize this section by the motivation for generation: synthesis and augmentation for utility, and privacy and trustworthiness for safe deployment.

\subsection{Synthesis and Augmentation}\label{sec:tabular-synthesis}

The first group of tabular papers uses diffusion models to generate synthetic financial records that can replace, supplement, or rebalance real training data. \citet{sattarov2023findiff} propose FinDiff for financial tabular data generation, targeting mixed financial records such as credit, payment, and fund-like tables. Entity-based financial tabular synthesis~\cite{liu2024enttabdiff} extends this idea to entity-structured records, where generated rows must remain plausible within the attributes and relationships of financial entities. \citet{sattarov2024imbfindiff} focus on class imbalance: their conditional diffusion model generates minority-class financial records so that downstream classifiers can learn rare but important events. 
Fraud-focused work follows the same rare-event logic: \citet{pushkarenko2024synthetic} study synthetic fraud-record generation for improving downstream fraud detection, where usefulness is measured by whether generated samples improve rare-event classification. Flow-based latent-space augmentation work~\cite{ihsan2025latentflow} follows the same utility-driven logic, using latent flow-based samples to improve prediction on credit and customer tabular tasks. \citet{kuo2026emdt} apply a diffusion transformer to fraud-detection tables, where the generated records are useful only if they improve rare-event detection without distorting the fraud/non-fraud boundary.

\subsection{Privacy and Trustworthiness}\label{sec:tabular-privacy}

The second group treats synthetic or generated tabular records as objects that must satisfy privacy, compliance, validity, or explanation constraints. Financial institutions often cannot share raw customer, transaction, or credit records, so synthetic data are valuable only if they retain utility while reducing memorization and disclosure risk. Federated diffusion with differential privacy addresses this setting by learning tabular generators across decentralized financial datasets while limiting information leakage from any participant~\cite{sattarov2024dpfedtabdiff}. Privacy-preserving mixed-type diffusion models make the same concern explicit for heterogeneous financial tables, where leakage may occur through rare categorical combinations as well as continuous outliers~\cite{sattarov2025dpfindiff}. Measuring privacy risks and tradeoffs in financial synthetic data generation shifts the emphasis from model design to evaluation: utility, fidelity, and disclosure risk must be reported together rather than optimized separately~\cite{zuo2026privacyrisks}. 
\citet{cardei2026constrained} address trustworthiness from a constraint-satisfaction perspective, proposing constrained tabular diffusion for finance so that generated records respect domain rules rather than only matching marginal distributions.
\citet{zhang2026tabular} propose Tabular Diffusion Counterfactual Explanation, a classifier-guided diffusion method for heterogeneous tabular data that uses a Gumbel-softmax-based reverse process to generate valid and diverse counterfactuals for credit-lending and related tabular decision tasks.

\section{Other Structured Financial Data}\label{sec:structured}

Not all financial objects are ordinary time series, LOB states, or tabular rows. Some data objects carry mathematical structure that must be preserved by the generator. Examples include correlation matrices, implied-volatility surfaces, yield curves, derivative paths, and risk-neutral pricing objects. These objects are structured because validity is partly defined by financial constraints: correlation matrices should be symmetric and positive semidefinite, volatility surfaces should respect no-arbitrage and cross-maturity consistency, yield curves should remain smooth and economically plausible, and derivative paths may need to respect payoff or risk-neutral dynamics. For this reason, evaluation in this section is less about generic sample realism and more about whether generated objects remain financially admissible.

Matrix, surface, and curve papers make this constraint issue explicit. \citet{kubiak2024corrddpm} apply DDPMs to financial correlation matrices, proposing both unconditional and conditional generators that reproduce empirical market structure and regime differences, with a case study showing how synthetic matrices can augment data for asset allocation and risk modeling. \citet{jin2025ivsurface} forecast implied-volatility surfaces with generative diffusion models, where the generated object is a surface over strike and maturity. Term-structure work uses diffusion models to generate yield curves across maturities, making the data object a cross-maturity curve whose quality depends on smoothness, level-slope-curvature structure, and economic plausibility~\cite{fukunishi2025termstructure}. Related fixed-income work extends structured generation to treasury futures: \citet{zhang2026tfcodit} use a conditional diffusion transformer to synthesize treasury-futures time series under user-specified conditions.

Derivative-path and pricing papers use diffusion models for valuation-oriented structured data. \citet{zhao2025exotic} apply conditional diffusion to exotic option valuation and counterparty games, where generated paths must remain consistent with payoff conditions and valuation objectives. \citet{gao2025sdepaths} generate solution paths of Markovian stochastic differential equations using diffusion models, placing the emphasis on path generation under specified stochastic dynamics rather than empirical market-data synthesis. \citet{tiwari2026riskneutral} modify DDPM reverse dynamics with a closed-form risk-neutral score shift, generating asset-price paths whose discounted prices satisfy the martingale condition and can be used to price European and path-dependent derivatives.

\section{Open Directions and Conclusion}\label{sec:conclusion}

Across the reviewed literature, the main pattern is that financial generation is becoming increasingly data-object specific, which motivates three open directions for the next phase of research: building finance-specific evaluation benchmarks, understanding how diffusion and flow models scale in realistic market settings, and moving beyond generation toward decision-making.

\textbf{Evaluation and benchmark.}
The most urgent open problem is benchmark and evaluation discipline. Diffusion models in image, video, and language generation became cumulative partly because the community built shared datasets, standard protocols, visible leaderboards, and increasingly demanding evaluation suites~\cite{deng2009imagenet, heusel2017gans, unterthiner2018towards, huang2024vbench, wang2018glue}. Finance does not yet have an equivalent. Many papers use proprietary data, incompatible horizons, different preprocessing choices, weak or non-overlapping baselines, and downstream tasks that are difficult to compare. As a result, the field risks producing many plausible demonstrations but little cumulative evidence. A useful research agenda is to build object-specific benchmark suites rather than relying on ad hoc realism checks. Image generation benefited from standardized scores such as Inception Score (IS)~\citet{salimans2016improved} and Fréchet Inception Distance (FID)~\cite{heusel2017gans}, which made distributional realism numerically comparable across papers. 


\textbf{Scaling up.}
A second open direction is scaling. In language modeling, scaling laws made progress more predictable by relating loss to model size, data, and compute~\cite{kaplan2020scaling}, while compute-optimal training showed that data and parameters must be scaled jointly rather than increasing model size alone~\cite{hoffmann2022training}. Recent work suggests that diffusion transformers exhibit similar power-law behavior, with pretraining loss scaling predictably with compute and correlating with generation quality~\cite{liang2024scaling}. Finance has not yet had an analogous scaling moment. Existing studies remain fragmented across proprietary datasets, asset classes, sampling frequencies, horizons, and evaluation protocols, making it unclear whether larger financial diffusion models improve because of scale, data quality, conditioning, or task design. A useful research agenda is therefore to study financial scaling laws directly: how model performance changes with more assets, longer histories, higher-frequency observations, richer conditioning variables, and larger compute budgets. Such scaling evidence would help move the field from isolated demonstrations toward benchmarked and compute-aware financial generators.

\textbf{Going beyond generation.}
Most existing diffusion-based finance studies evaluate whether generated samples resemble real financial data or improve downstream tasks through scenario augmentation. A more ambitious direction is to use diffusion models directly as decision-makers. In the broader machine-learning literature, diffusion models have been used to generate behavior trajectories for planning, conditional action sequences, and reinforcement-learning policies~\cite{zhu2023diffusion, xu2025diffusion}. Finance has natural counterparts, including portfolio construction, hedging, execution, and market making, where the output is not only a plausible market path but an action under uncertainty. Future work could therefore move from asking whether generated samples look realistic to whether diffusion-based policies support better decisions under market uncertainty, risk limits, and regulatory constraints.

This survey argues for a data-first view of diffusion models in finance. Progress will not come from treating finance as another generic data modality or from comparing model families in isolation. It will come from building benchmarked, scaled, and constraint-aware generators whose validity is defined by the financial object they model. The next phase of diffusion models in finance should therefore move beyond asking whether diffusion can generate financial data, and instead ask whether such models can become trustworthy financial simulators and decision engines.

\bibliographystyle{ACM-Reference-Format}
\bibliography{references}

\end{document}